# Phone-sized whispering-gallery microresonator sensing system

Xiangyi Xu, Xuefeng Jiang, Guangming Zhao, and Lan Yang[*]

*Department of Electrical and Systems Engineering, Washington University, St Louis, Missouri 63130, USA*
[*]*yang@seas.wustl.edu*

**Abstract:** We develop a compact whispering-gallery-mode (WGM) sensing system by integrating multiple components, including a tunable laser, a temperature controller, a function generator, an oscilloscope, a photodiode detector, and a testing computer, into a phone-sized embedded system. We demonstrate a thermal sensing experiment by using this portable system. Such a system successfully eliminates bulky measurement equipment required for characterizing optical resonators and will open up new avenues for practical sensing applications by using ultra-high $Q$ WGM resonators.

## 1. Introduction

In the past two decades, whispering-gallery-mode (WGM) microresonators have found broad applications in photonics [1], such as cavity quantum electrodynamics [2], optomechanics [3-8], bio/chemical sensors [9-12], nonlinear optics [13-17] and microlasers [18-20], which is attributed to significantly enhanced light-matter interaction rising from ultra-high quality ($Q$) factors and small mode volumes. However, till now all the demonstrations were implemented in optical laboratories with well-set equipment on bulky optical tables, which limit the practical applications of WGM microresonators, *e.g.*, various kinds of sensing, including sensing of single nanoparticle [4-7], biomolecule [21,22], magnetic field [23,24], angular velocity [25-27], gas [28,29], *etc*. The obstacles of practical applications for WGM sensors lie on two factors: i) the challenge of long-term stability for tapered fiber coupling of cavity modes outside the laboratory, and ii) bulky commercial equipment needed for testing cavity modes, including not only a laser source and a detector but also a function generator and an oscilloscope. The first challenge has been partly solved by packaging the microresonators with the tapered fiber waveguide in a low refractive index polymer matrix [30-32], which can not only isolate the coupling from the environmental perturbation but also achieve relative high $Q$ factor and coupling efficiency. As for the second challenge, however, no progress has been made so far because of the difficulty in combing all those commercial instruments into a small portable circuit board.

Here we report the first realization of a compact WGM sensing system, which integrates a tunable laser, a current source, a temperature controller, a function generator, an oscilloscope, a photodiode detector, a testing computer with the customized testing software, and a packaged WGM sensor into a phone-sized embedded system. Utilizing such a compact system, we demonstrate a thermal sensing experiment. This portable measurement system opens up new avenues for practical sensing applications by using ultrahigh-$Q$ WGM resonators.

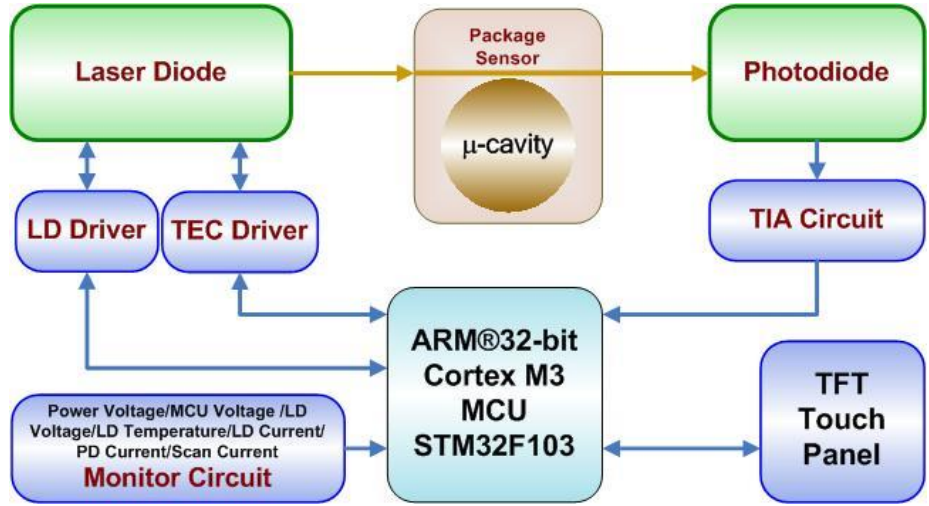

Fig. 1. Schematic diagram of a WGM based portable sensor system.

## 2. The architecture and principle of the portable WGM measurement system

As showed in Fig. 1, the portable system consists of a Distributed Bragg Reflector (DBR) laser, a laser diode (LD) driver, a thermo-electric cooler (TEC) driver, a photodetector, a home-made transimpedance amplifier (TIA) circuit, a monitor circuit, a thin-film transistor (TFT) touch panel and a microcontroller unit (MCU) ARM Cortex-m3 processor. The single-mode tunable DBR laser with the central wavelength of around 976 nm, the linewidth of 10 MHz, and the output power of 35 mW, is utilized as a probe laser source in the system. Specifically, the fiber pigtailed DBR laser in a 14 pin butterfly package includes a TEC, a thermistor, and a monitor photodiode. Laser current driver and TEC driver with 0.9 mK temperature stability are used for stabilizing of the laser frequency, both of which can be exploited for modulating or scanning the probe laser wavelength.

The ARM Cortex-m3 processor serves as the brain of the whole portable WGM measurement system, whose primary function is running the embedded uC/OS-II operating system and controlling the laser current driver and the temperature controller. Its digital-to-analog converter (DAC) and analog-to-digital converter (ADC) interfaces play the roles of function generator and oscilloscope, respectively. In the experiments, the transmission spectrum of the packaged WGM sensor is detected by the photodetector. A customized TIA circuit was designed to convert small current output from the photodetector to voltage with appropriate gain, which was connected to the processor's ADC interface to deliver a normalized transmission spectrum in the front panel of the portable system as shown in Fig. 2. Also, a particular monitor circuit and embedded software program were also developed to monitor the key parameters of the system, such as power supply voltage, the processor voltage, LD voltage, LD current, LD temperature, *etc*.

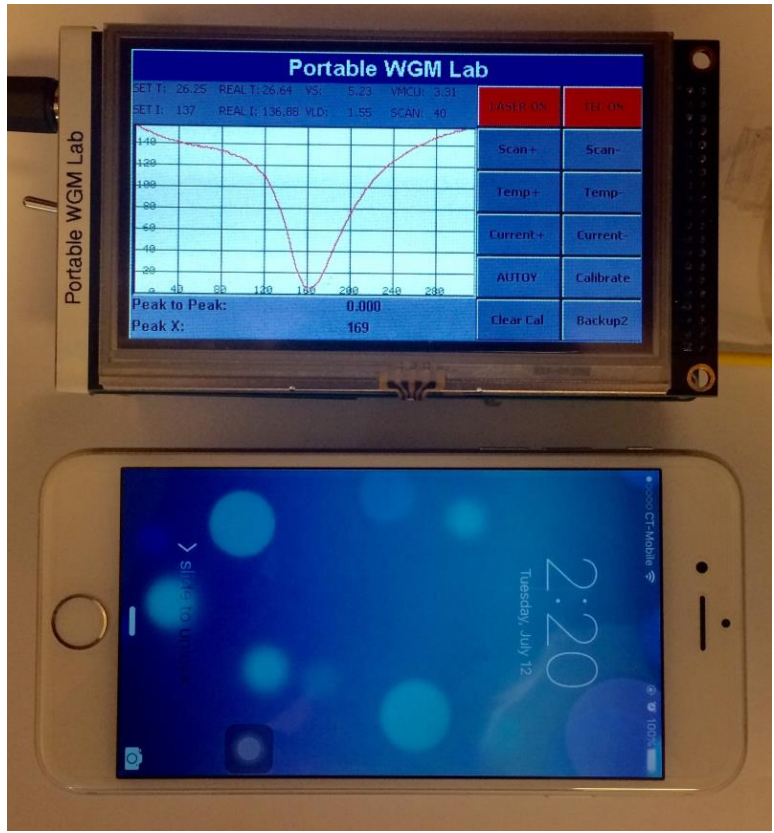

Fig. 2. A photograph of the portable WGM testing system compared with a similar-sized iPhone 6s.

The key monitoring parameters of the measurement system, such as the laser diode current/temperature, the voltage of power supply, the laser current scannig range, *etc.*, are displayed on the TFT touchscreen (Fig. 2). Besides, the key functions to control the measurement, such as tuning and scanning the laser frequency around a resonance, can be carried out through the same TFT touchscreen, which covers the central controlling board of the portable system, as showed in Fig. 2. In addition, some of the data analyzing processes could also be displayed on the touchscreen in real time, which is achieved by the mathematic algorithm integrated into the embedded software for the ARM processor. More complicated data processing can be carried out via a USB interface by transferring the data to a computer. Figure 2 shows a photograph of the portable WGM testing system compared with an iPhone 6s, where a calibrated WGM transmission spectrum, as well as all the monitoring and controlling parameters, are displayed.

## 3. Experimental characterization of the portable system

Frequency scanning range up to 24 GHz could be achieved by sweeping the LD current with a tuning coefficient of 0.002 nm/mA. Specifically, the frequency scanning is conducted by applying a 40 mA amplitude sawtooth wave to the laser current at a fixed current and a fixed TEC temperature. In addition, the central frequency of the scanning process can be adjusted by changing the injection current or TEC temperature with a tuning coefficient of 0.07 nm/$^o$C. Meanwhile, the optical power of different frequency is monitored by the integrated monitor photodiode and displayed on the TFT screen in real time. Therefore, a typical transmission spectrum of a packaged WGM resonator can be collected by using the portable system, as shown in Fig. 3(a).

Since the frequency scanning is performed by sweeping the laser current, the linearity of the frequency detuning as a function of current change is critical for the experiment. We further investigated the central resonant frequency detuning of a packaged cavity mode by changing the laser current. Figure 3(b) shows a perfect linear fitting of frequency detuning as a function of the injection current in the laser diode, which validates the linear frequency scanning controlled by the current. On the other hand, the probe laser frequency can also be tuned by adjusting the LD temperature through TEC driver, and thus we performed the similar experiment by changing the temperature, as shown in Fig. 3(c). It's worth noting that the linearity of this fitting is not so good as that in the case where current is adjusted to tune the frequency, which can be attributed to the intrinsic characteristic of temperature tuning of DBR laser as well as the uncertainty in the temperature measurement. Thus, the temperature tuning method is only used for changing the central frequency instead for frequency scanning in the experiment.

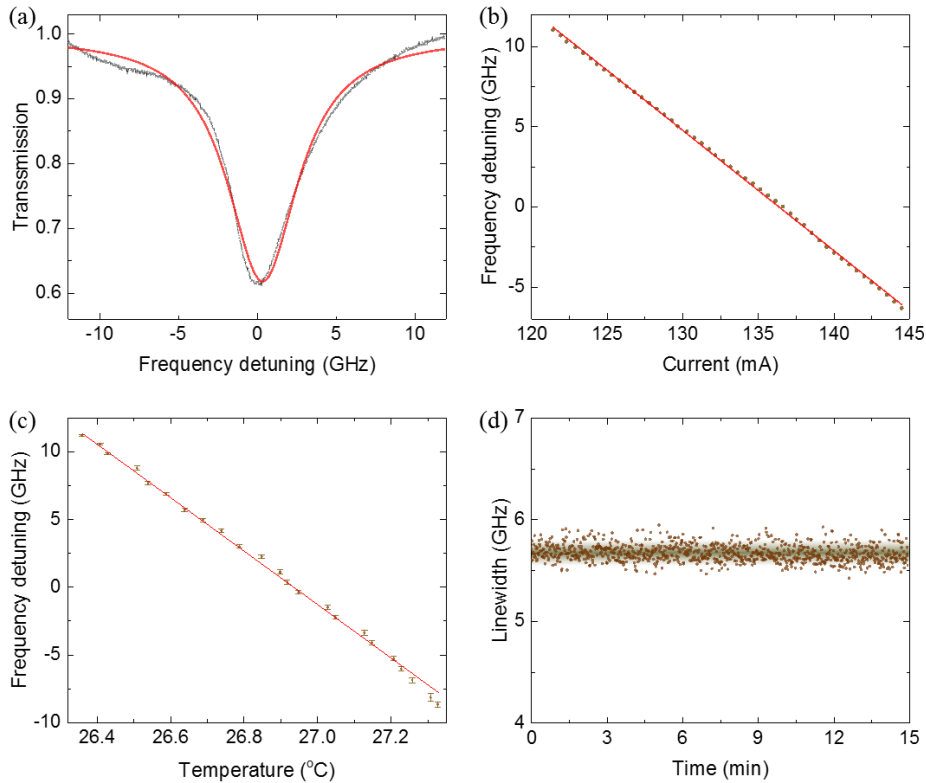

Fig. 3 (a) A typical transmission spectrum of a packaged WGM resonator monitored by the portable WGM testing system. (b) Resonant frequency detuning as a function of the injection current in the laser diode. (c) Resonant frequency detuning as a function of the temperature of the laser diode. (d) The time stability of the linewidth of a resonance in the packaged resonator monitored by the portable WGM testing system.

To study another important parameter, *i.e.*, the stability of the system, we investigated the time trace of the linewidth of a resonant mode in a packaged resonator. As shown in Fig. 3(d), the average linewidth is 5.66 GHz with an uncertainty of 90 MHz. Note that the uncertainty of the cavity mode is about 1.6% of the corresponding linewidth here, which is similar to the results of laboratory's setup (1.3%) [11]. It clearly demonstrates that this portable WGM testing system has the similar performance with the measurement system consisting of bulky commercial equipment used in the lab.

**4. Thermal sensing experiment**

As a sensing application of this portable testing system, we further performed a thermal sensing experiment by using a microtoroid on a silicon chip [33-35]. In this experiment, we placed another TEC under the microtoroid to change the local temperature by adjusting the TEC driver current. The resonant frequency shifts of a particular WGM around 976 nm were then recorded by the portable system as a function of temperatures, as shown in Fig. 4. A total frequency detuning of 21.9 GHz as a result of 8.8 ℃ temperature change was observed with a sensitivity of 2.49 GHz/K, *i.e.*, 8.13 pm/K., which agrees well with the previous silica based WGM thermal sensor [36,37]. The similar sensitivity of the thermal sensing experiment further proves that this portable WGM testing system is a promising candidate to replace the bulky equipment in the lab for WGM experiments, which will enhance the impact of WGM resonators to a variety of practical sensing applications.

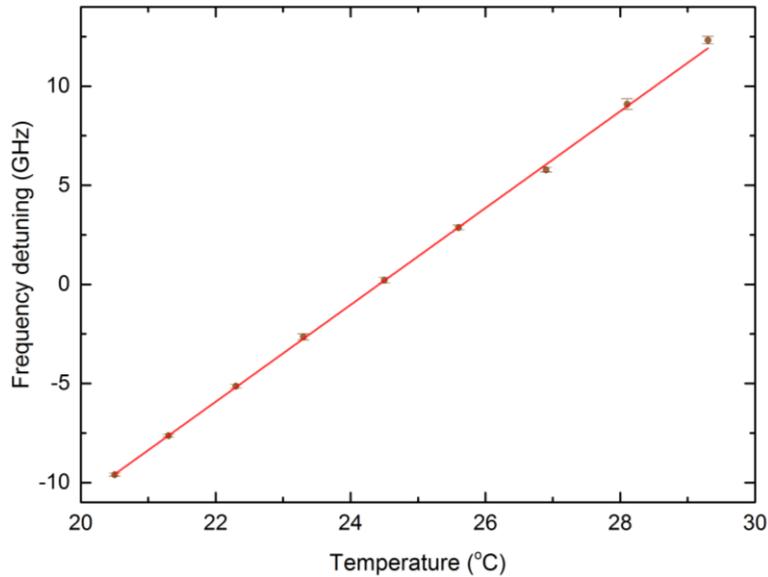

Fig. 4. Thermal sensing experiment of a normal microtoroid by using the portable system.

**5. Conclusions**

In summary, we have demonstrated a compact WGM testing system for sensing by integrating all the essential functions, previously provided by bulky equipment, into a phone-sized embedded system. By scanning the laser current, a cavity mode with the measurement stability similar to that observed in the traditional WGM testing system in lab setting was presented. Furthermore, we demonstrate the thermal sensing experiment of a microtoroid by using this portable system with a sensitivity of 8.13 pm/K, matching well with the result in literature. This portable WGM testing system opens up new avenues for a variety of sensing applications by using ultra-high-$Q$ WGM resonators, and represents a milestone for practical applications of WGM resonator and other resonant structures.

**Acknowledgments**

X. Xu and X. Jiang contributed equally to this work. The authors thank Dr. Xuan Zhang for helpful discussion. This work was supported by ARO grant No. W911NF-12-1-0026.